# On the relativistic generalization of Maxwell's velocity distribution

Jian-Miin Liu*
Department of Physics, Nanjing University
Nanjing, The People's Republic of China
*On leave. E-mail address: liu@phys.uri.edu

ABSTRACT
Some problems relevant to the relativistic generalization of Maxwell's velocity distribution are discussed.

## 1

Maxwell read his paper on equilibrium velocity distribution before the British Association in Aberdeen on September 21, 1859. In this paper [1], based on two assumptions (1) the velocity distribution function is spherically symmetric and (2) The x-, y- and z- components of velocity are statistically independent, Maxwell derived his equilibrium velocity distribution,

$$M(y^1,y^2,y^3)dy^1dy^2dy^3 = N\left(\frac{m}{2\pi K_B T}\right)^{3/2} \exp\left[-\frac{m}{2K_B T}(y)^2\right] dy^1 dy^2 dy^3, \tag{1a}$$

$$M(y)dy = 4\pi N\left(\frac{m}{2\pi K_B T}\right)^{3/2} (y)^2 \exp\left[-\frac{m}{2K_B T}(y)^2\right] dy, \tag{1b}$$

where $y^r = dx^r/dt$, $r=1,2,3$, is the well-defined Newtonian velocity in the usual inertial coordinate system $\{x^r, t\}$, $y=(y^r y^r)^{1/2}$, N is the number of particles, m their rest mass, T the temperature and $K_B$ the Boltzmann constant. Assumptions (1) and (2) reflect structural characteristics of the velocity space,

$$dY^2 = \delta_{rs} dy^r dy^s, \quad r,s=1,2,3, \tag{2}$$

in non-relativistic mechanics. Maxwell's velocity distribution can be directly deduced from the Boltzmann distribution in the one-particle phase space in non-relativistic statistical mechanics.

Maxwell's velocity distribution is non-relativistic. From the theoretical point of view, as special relativity is a part of the laws of Nature, it is natural to consider the relativistic generalization of Maxwell's velocity distribution. From the practical point of view, the relativistic equilibrium velocity distribution is necessary when we deal with statistical calculations where most particles crowd in the high-energy region or most particles crowd in the low-energy region but these particles in the low-energy region are not involved in the dealt statistical calculations. Calculations of the nuclear fusion reaction rate are an example of this type.

## 2

In our recent work [2,3], we made the relativistic generalization of Maxwell's velocity distribution through analyzing relativistic velocity space,

$$dY^2 = H_{rs}(y) dy^r dy^s, \quad r,s=1,2,3, \tag{3a}$$

$$H_{rs}(y) = c^2 \delta^{rs}/(c^2-y^2) + c^2 y^r y^s/(c^2-y^2)^2, \text{ real } y^r \text{ and } y<c, \tag{3b}$$

in the usual velocity-coordinates $\{y^r\}$, $r=1,2,3$, where c is the speed of light. This velocity space can be represented by

$$dY^2 = \delta_{rs} dy'^r dy'^s, \quad r,s=1,2,3, \tag{4}$$

in the so-called primed velocity-coordinates $\{y'^r\}$, $r=1,2,3$, where

$$dy'^r = A^r_s(y) dy^s, \quad r,s=1,2,3, \tag{5a}$$

$$A^r_s(y) = \gamma \delta^{rs} + \gamma(\gamma-1) y^r y^s / y^2, \tag{5b}$$

with

$$\gamma = 1/(1-y^2/c^2)^{1/2}, \tag{6}$$

because $\delta_{rs} A^r_p(y) A^s_q(y) = H_{pq}(y)$. We call $y'^r$, $r=1,2,3$, the primed velocity.

Using calculation techniques in Riemann geometry, we can find

$$Y^2 = \delta_{rs} y'^r y'^s, \tag{7a}$$

$$Y^2=[\frac{c}{2y}\ell n\frac{c+y}{c-y}]^2\delta_{rs}dy^r dy^s, \quad (7b)$$

$$y'^r=[\frac{c}{2y}\ell n\frac{c+y}{c-y}]y^r, \; r=1,2,3, \quad (8a)$$

$$y'=\frac{c}{2}\ell n\frac{c+y}{c-y}, \quad (8b)$$

where $(y'^1, y'^2, y'^3)$ and $(y^1, y^2, y^3)$ denote the same point in the relativistic velocity space, $y'=(y'^r y'^r)^{1/2}$. The relativistic velocity space is characterized by a finite boundary at $y=c$ and the Einstein addition law in the usual velocity-coordinates and also characterized by unboundedness and the Galilean addition law in the primed velocity-coordinates. Differentiating Eq. (8b) immediately yields

$$dy'=dy/(1-y^2/c^2). \quad (9)$$

The Euclidean structure of the relativistic velocity space in the primed velocity-coordinates convinces us of Maxwell's equilibrium distribution of primed velocities,

$$P(y'^1,y'^2,y'^3)dy'^1 dy'^2 dy'^3 = N(\frac{m}{2\pi K_B T})^{3/2}\exp[-\frac{m}{2K_B T}(y')^2]dy'^1 dy'^2 dy'^3, \quad (10a)$$

$$P(y')dy'=4\pi N(\frac{m}{2\pi K_B T})^{3/2}(y')^2\exp[-\frac{m}{2K_B T}(y')^2]dy'. \quad (10b)$$

Inserting Eqs.(5a-5b), (8b) and (9) into Eqs.(10a-10b), we obtain the relativistic equilibrium distribution of Newtonian velocities,

$$P(y^1,y^2,y^3)dy^1 dy^2 dy^3 = N\frac{(m/2\pi K_B T)^{3/2}}{(1-y^2/c^2)^2}\exp[-\frac{mc^2}{8K_B T}(\ell n\frac{c+y}{c-y})^2]dy^1 dy^2 dy^3, \quad (11a)$$

$$P(y)dy=\pi c^2 N\frac{(m/2\pi K_B T)^{3/2}}{(1-y^2/c^2)}(\ell n\frac{c+y}{c-y})^2\exp[-\frac{mc^2}{8K_B T}(\ell n\frac{c+y}{c-y})^2]dy. \quad (11b)$$

This distribution fits to the Maxwellian distribution for low-energy particles ($y<<c$) but substantially differs from the Maxwellian distribution for high-energy particles. As y goes to c, it falls to zero slower than any exponential decay and faster than any power-law decay [4]. Such a new kind of decay mode has been used to explain the observed non-Maxwellian high-energy tails in velocity distributions of astrophysical plasma particles [4]. The distribution Eqs.(11a-11b) has been also used to calculate the nuclear fusion reaction rate [5]. The nuclear fusion reaction rate based on the relativistic equilibrium distribution, R, has a reduction factor with respect to that based on the Maxwellian velocity distribution, $R_M$,

$$R=\frac{\tanh Q}{Q}R_M, \quad (12)$$

where Q depends on the temperature, reduced mass and atomic numbers of the studied nuclear fusion reactions. Since $0<Q<\infty$, $0<\tanh Q/Q<1$ and

$$0<R<R_M. \quad (13)$$

Eqs.(12) and (13) signify much to the solar neutrino problem.

Our paper [2] was submitted to physics-professional journals for publication but refused. From referees, we got a gain to knowledge about other trials of relativistic generalization of Maxwell's velocity distribution.

In a referee's report (Ref. X/13559/XXX), we read: "The velocity as contrasted to the momentum distribution for classical relativistic particles is well known: N.E.Frankel, K.C.Hines, R.L.Dewar---Energy loss due to binary collisions in a relativistic plasma, Phys.Rev. A20, 2120-2129 (1979)" We went for the mentioned paper but was unsuccessful. For the relativistic generalization of Maxwell's velocity distribution, there is nothing in the paper except to write down velocity distribution formula

$$f(w)dw \sim w^2 \left(\frac{1}{\sqrt{1-w^2/c^2}}\right)^5 \exp\left\{-\frac{mc^2}{K_B T\sqrt{1-w^2/c^2}}\right\} dw,$$

(we omitted a proportional coefficient) and to name it the isotropic relativistic Maxwellian velocity distribution. There is no explanation, no derivation and no reference for this formula. In f(w), exponential factor $\exp\{-mc^2/K_B T(1-w^2/c^2)^{1/2}\}$ seems to come from substituting the relativistic one-particle energy for the non-relativistic one in Maxwell's velocity distribution, while factor $w^2(1-w^2/c^2)^{-5/2}$ is quite out of understanding.

In another report (Ref. XXXX/142/2002), a referee wrote: "however, the simple relativistic generalization of Maxwell's distribution function is just:

$$f(p) = \exp\left\{-\frac{1}{2K_B T}[(p^2c^2+m^2c^4)^{1/2} - mc^2]\right\}$$

i.e. the replacement of the non-relativistic kinetic energy by its relativistic equivalent." Correcting an error in the exponent of f(p) and representing f(p) in terms of velocity, we find referee's velocity distribution proportional to

$$f(y)dy^1 dy^2 dy^3 \sim \exp\left\{-\frac{1}{K_B T}\left[\frac{mc^2}{\sqrt{1-y^2/c^2}} - mc^2\right]\right\} dy^1 dy^2 dy^3.$$

The referee also recognizes substituting the relativistic one-particle kinetic energy for the non-relativistic one in Maxwell's velocity distribution.

Maxwell's velocity distribution Eq.(1a) can be rewritten as

$\sim \exp\{-E/K_B T\} dy^1 dy^2 dy^3,$ \hfill (14a)
$E = m(y)^2/2.$ \hfill (14b)

Two referees, from two different journals, both suggested the replacement of the non-relativistic kinetic energy by its relativistic equivalent, in Maxwell's velocity distribution, for relativistic generalizing Maxwell's velocity distribution. They asked

$\sim \exp\{-E/K_B T\} dy^1 dy^2 dy^3,$ \hfill (15a)
$$E = \frac{mc^2}{\sqrt{1-y^2/c^2}} \text{ or } \left[\frac{mc^2}{\sqrt{1-y^2/c^2}} - mc^2\right],$$ \hfill (15b)

as a relativistic generalization of Maxwell's velocity distribution, where Eq.(15a) is the same as Eq.(14a). However, Eq.(14a) is linked to the Boltzmann distribution in the one-particle phase space, in non-relativistic statistical mechanics, for a confined many-particle system with weak interactions and zero external forces in volume V,

$\sim \exp\{-E/K_B T\} d\omega,$ \hfill (16)

where $d\omega = dx^1 dx^2 dx^3 dp^1 dp^2 dp^3$ is a volume element of the phase space, $p^r = my^r$, r=1,2,3, while Eq.(15a) does not exist in the framework of special relativity.

In non-relativistic statistical mechanics, one can depend upon Boltzmann's hypothesis on the equality of the probability of equal volume element to derive the Boltzmann distribution. It should be noted that Boltzmann's hypothesis is appropriate only for uniform phase space structures. The phase space is a direct product of coordinate space and momentum space. In non-relativistic mechanics, coordinate space is $dX^2 = \delta_{rs} dx^r dx^s$, r=1,2,3, while momentum space is $dP^2 = \delta_{rs} dp^r dp^s$ due to velocity space Eq.(2), where $p^r = my^r$, r=1,2,3, and P=mY. In the framework of non-relativistic mechanics, the one-particle phase space has a uniform structure in $\{x^r; p^r\}$, r=1,2,3. On the contrary, using the relativistic relations between momentum and velocity,

$p^r = my^r/(1-y^2/c^2)^{1/2}$, r=1,2,3, \hfill (17a)
$y = cp/\sqrt{p^2 + m^2c^2}$, \hfill (17b)

we find from Eq.(7b),

$$P^2 = \left[\frac{mc}{2p} \ln \frac{(m^2c^2+p^2)^{1/2}+p}{(m^2c^2+p^2)^{1/2}-p}\right]^2 \delta_{rs} p^r p^s, \text{ r,s=1,2,3,}$$ \hfill (18)

where $p^2=\delta_{rs}p^r p^s$, $r,s=1,2,3$. The relativistic momentum space can not be of uniform structure in $\{p^r\}$. The relativistic one-particle phase space can not be of uniform structure in $\{x^r; p^r\}$, either. In the framework of special relativity, we do not have Eq.(16). In the framework of special relativity, we do not have Eq.(15a), either, even for a confined many-particle system with weak interactions and zero external forces in volume V. For the relativistic generalization of Maxwell's velocity distribution, it is non-sense to put Eq.(15b) in Eq.(15a) or to replace the non-relativistic kinetic energy by its relativistic equivalent in Maxwell's velocity distribution.

ACKNOWLEDGMENT

The author greatly appreciates the teachings of Prof. Wo-Te Shen. The author thanks Dr. P. P. Rucker for helpful suggestions.